\documentstyle[12pt]{article}
  
\def\p{\partial}
\def\be{\begin{equation}}
\def\ee{\end{equation}}
\def\ben{\begin{displaymath}}
\def\een{\end{displaymath}}
\def\ba{\begin{array}{c}}
\def\ea{\end{array}}
\begin{document}

\titlepage
\vspace*{2cm}

\begin{center}{\large \bf

Linear representation of energy-dependent Hamiltonians

}\end{center}

\vspace{10mm}

\begin{center}
Miloslav Znojil

 \vspace{3mm}

\'{U}stav jadern\'e fyziky AV \v{C}R, 250 68 \v{R}e\v{z},
Czech Republic\\

e-mail: znojil@ujf.cas.cz

\end{center}

\vspace{5mm}

\section*{Abstract}

Quantum mechanics abounds in models with Hamiltonian operators
which are energy-dependent. A linearization of the underlying
Schr\"{o}dinger equation with $H = H(E)$ is proposed here via an
introduction of a doublet of separate energy-independent
representatives $K$ and $L$ of the respective right and left
action of $H(E)$. Both these new operators are non-Hermitian so
that our formalism admits a natural extension to non-Hermitian
initial $H(E)$s. Its applicability may range from pragmatic
phenomenology and variational calculations (where all the
subspace-projected effective operators depend on energy by
construction) up to perturbation theory and quasi-exact
constructions.

\vspace{5mm}

PACS 02.30.Tb  03.65Ca  03.65.Db  03.65.Ge

\newpage

\section{Introduction}

In non-relativistic quantum mechanics the role of time is
exceptional as it remains, up to a few really exotic models
\cite{clock}, unquantized. This makes the current Schr\"{o}dinger
equation $i\,\hbar\,\p_t \psi(t) = H\,\psi(t)$ easily integrated
in all the time-independent cases where we get $\psi(t)= \exp
[i\,(t-t_0)\,H]\,\psi(t_0)$. This means that we just have to solve
the time-independent Schr\"{o}dinger equation
 \be
 H\,|\Psi^{(n)}\rangle =E^{(n)}\,|\Psi^{(n)}\rangle\,
 \label{SE0}
 \ee
and evaluate $\psi(t)$ using the spectral representation of the
Hamiltonian operator,
 \be
 H= \sum_n\,|\Psi^{(n)}\rangle \,E^{(n)}\,\langle\Psi^{(n)}|\,.
 \label{SE1}
 \ee
The picture becomes less transparent in all the time-dependent
regimes where we have to deal, e.g., with the time-dependent
boundary conditions \cite{Seba} and where the definition of
$H=H(t)$ becomes transformed and transferred to $H = H(E)$ in an
appropriate Fourier-transformation modification of eq.
(\ref{SE0}),
 \be
 H\left (E^{(n)}\right )\,|\Psi^{(n)}\rangle
  =E^{(n)}\,|\Psi^{(n)}\rangle\,.
 \label{SE}
 \ee
In our present short note we intend to make a few comments
concerning the latter ``non-linear" Schr\"{o}dinger equation.

\section{Motivation \label{motivate}}

\subsection{An emergence of the energy-dependence in physics}

The energy-dependent version (\ref{SE}) of the current bound-state
Schr\"{o}dinger equation (\ref{SE0}) is far from being an exotic
problem without any practical relevance. Almost an opposite
statement is true, at least in principle. Indeed, whenever we
contemplate a ``real" system and construct its manifestly
energy-independent ``realistic" Hamiltonian $H_R$ with due care,
we must keep trace of many ``possibly relevant" degrees of
freedom. In the next step we usually decide, on some more or less
intuitive grounds, that the ``really relevant" part of $H_R$ might
be just its projection $H_{(MS)}=P\,H_R\,P$ on a certain ``model"
Hilbert subspace.

Having finished such a procedure (which is quite common, say, in
nuclear physics \cite{Feshbach}) one is obliged to compare the
spectra $\{E_R\}$ (of $H_R$) and $\{E_{(MS)}\}$ (of $H_{(MS)}$)
and to verify, in this way, the quality and/or reliability of our
initial intuition. A formal tool is provided by perturbation
theory \cite{Kumar} in terms of which we may arrive at a more
quantitative verification of our hypotheses by the almost
elementary replacement of the ``exact" equation $
H_R\,|\Psi_R\rangle =E_R\,|\Psi_R\rangle$ by its two projections
$P$ and $Q=1-P$,
 \begin{eqnarray*}
 P\,\left (H_R-E_R\right )\,P\,|\Psi_R\rangle +
 P\,\left (H_R-E_R\right )\,Q\,|\Psi_R\rangle =0,
\\
 Q\,\left (H_R-E_R\right )\,P\,|\Psi_R\rangle +
 Q\,\left (H_R-E_R\right )\,Q\,|\Psi_R\rangle =0\,.
 \end{eqnarray*}
Eliminating the irrelevant projection $Q\,|\Psi_R\rangle$ we
easily arrive at the {\em strictly equivalent} reduced eigenvalue
problem $H_{(red)}(E_R) \,|\varphi\rangle =E_R\,|\varphi\rangle$
of the form (\ref{SE}) with
 \be
 H_{(red)}(E)=
 \left \{
 H_R+
 H_R\,Q\,
 \left [ \frac{Q}{Q\,\left (H_R-E \right )\,Q}
  \right ] \, Q\,H_R
 \right \}\,.
 \label{exact}
 \ee
Using such a construction of $H_{(red)}=H_{(red)}(E)$, all the
energy errors vanish and we have just the identical energy spectra
$\{E_R\} \equiv \{E_{(red)}\}$, in spite of the {\em arbitrary}
reduction $|\Psi_R\rangle \to P\,|\Psi_R\rangle \equiv
|\varphi\rangle $ of the Hilbert space itself.

\subsection{Energy-dependent models in computations}

The role of the above exact formula (\ref{exact}) in practical
computations is indirect because the energy-dependent operator
difference between the exact $H_R$ and reduced $H_{(red)}(E_R)$
is, undoubtedly, hardly obtainable in closed form. This means that
the unique and exact parameter-dependence of $H_{(red)}(z)$ is too
complicated. In practical computations it invariably requires a
perturbative, variational or at least empirical or trial-and-error
simplification $H_{(red)}(z) \to H(z)$.  At all the relevant real
values of $z$ this re-opens the possibility of employing the
spectral decomposition,
 \be
 H_{}(z)= \sum_n\,|\Psi^{(n)}_{}(z)\rangle \,E^{(n)}_{}(z)
 \,\langle\Psi^{(n)}_{}(z)|\,, \ \ \ \ \ \ \ \ \  \
  z \in \left\{ E_R^{(m)} \right \}_{m=0}^\infty\,.
 \label{SE2}
 \ee
We see that the work with the energy-dependent Hamiltonians may be
both feasible and useful \cite{ZBJ}. One only has to keep in mind
that it still requires comparatively lengthy calculations because
at any pre-selected superscript we must guarantee that the value
of $z$ satisfies the self-consistency condition
 \be
 z_{phys} =E^{(n)}_{}(z_{phys})\,.
 \label{nelin}
 \ee
This means that the algorithm of construction of bound states
generated by any energy-dependent Hamiltonian $H_{}(E)$ remains
non-linear, with all the unpleasant mathematical consequences
including the possible non-existence or redundancy of the real
(and even complex) solutions of eq. (\ref{nelin}) at every $n$.

\section{Examples of applicability \label{protoHilbert} }

\subsection{Energy scale and its subdomains}

\subsubsection{Thresholds in phenomenological physics}

In the majority of phenomenological models $H_{ph.}$, one does not
specify any upper limit $E_{max}$ of their applicability. Thus,
atomic and nuclear systems are very often described by
non-relativistic Schr\"{o}dinger equations with interactions
reduced to a local two-body phenomenological potential
$V=V(x_i-x_j)$ \cite{he3}. Of course, the credibility of such a
model becomes very low at the higher energies so that we tacitly
have to assume the existence of a complete, two-component
description
 \be
 \begin{array}{cc}
 H = H_{low\,en.}&  E\leq E_{crit.}\\
 H = H_{high\,en.}&  E > E_{crit.}
\ea
 \label{mostele}
 \ee
where the precise localization of the onset of the new physics
(expected to occur somewhere near the ``threshold" energy $
E_{crit.} \approx 300 MeV$ in nuclear physics) is rarely
considered really important because, for years, the detailed
structure of $H_{high\,en.}$ has been believed to be prohibitively
complicated. Still, due to the steady increase of our computing
power, the high-energy parts of eq. (\ref{mostele}) and of its
step-function energy-dependence generalizations
 \be
 H = H_{a,b}\ {\rm for}\  E \in (E_a,E_b)
 \label{stele}
 \ee
might acquire the status of a tractable problem quite soon
\cite{ZBJ,my}.

\subsubsection{Smoothly $E-$dependent Hamiltonians }

One of the best known theoretical sources of differences between
the respective low- and high-energy Hamiltonians are relativistic
phenomena (e.g., the decays and the emergence of various new
degrees of freedom) and corrections (say, in the form of a weakly
energy-dependent effective mass in nuclear \cite{Lichard} and
particle \cite{Licharda} physics). In such a setting, the building
of a model should start in the relativistic kinematical regime and
proceed towards the lower energies by a systematic simplification
of $H_{high\,en.}$. One usually works with a well defined
energy-dependence in $H_{high\,en.}(E)$ which is smooth. A wealth
of explicit examples of this type fits in the scheme of eq.
(\ref{stele}) and occurs in numerous applications \cite{Hirota}.

Naturally, the study of the models $H(E)$ with a smooth energy
dependence attracts attention also by its methodic aspects. To
grasp their flavor, the reader is recommended to play with the
harmonic-oscillator example of ref. \cite{my},
 \be
 -\,\frac{\hbar^2}{2m(E^{(n)})}\ \frac{\rm d^2}{\rm dx^2}\,
  \Psi^{(n)}({x}) +
 g\,x^2\,
 \Psi^{(n)}({x})= E^{(n)} \Psi^{(n)}({x})\,,
 \label{hSE}\
 \ee
an extremely easy solution of which illustrates many paradoxes
exhibited by much more complicated realistic models.

\subsection{Energy-dependence as a formal freedom}

\subsubsection{Non-linearity and non-orthogonality}

The key constraint (\ref{nelin}) may be satisfied at a large as
well as empty set of real roots at a given $H(E)$ and $n$,
 \ben
 z_{phys} =E^{(n,1)},E^{(n,2)},\ldots,   E^{(n,m(n))}\,.
 \een
In a compactified notation using multi-indices $\alpha=(n,j)$ we
shall abbreviate $E^{(n,j)}=E_\alpha$ and denote
$|\Psi^{(n)}_{}(z_{phys})\rangle = |\phi_\alpha \rangle$ in order
to identify characteristic difficulties which arise in connection
with the energy dependence in $H = H(E)$.

First of all, we have to emphasize that the standard orthogonality
relations between the separate bound states become lost. Even
though the norm  $ ||\phi_\alpha||$ of each particular eigenstate
may easily be fixed through evaluation of the self-overlaps
$\langle \phi_\alpha |\phi_\alpha \rangle $, it is necessary to
evaluate also all the off-diagonal overlaps
 \be
 \langle \phi_\alpha |\phi_\beta \rangle =R_{\alpha, \beta}
 \label{overlaps}
 \ee
which need not non-vanish in general. Secondly, whenever we have
to deal with an energy-dependent model where the energies remain
discrete, we may alter the denotation of $H=H(E_\alpha)
=H(E^{(n,j)})= H^{(n)}(E^{(n,j)})$. In particular, the latter
convention may prove suitable whenever the partial
energy-independence property
 \be
 H^{(n)}(E^{(n,j)}) =
 H^{(n)}(E^{(n,k)}), \ \ \ \ \ \ j, k = 1, 2, \ldots, m(n)\,
 \label{notace}
 \ee
is encountered as a generalization of the step-shaped energy
dependence~(\ref{stele}).

\subsubsection{Quasi-exact solvability }

We shall see below that one of the most important technical
assumptions of the feasibility of the work with any $H=H(E)$ is
the feasibility of the evaluation of the overlap matrix $R$
(\ref{overlaps}). This is only trivial in the completely solvable
and fully energy-independent limit of eq. (\ref{hSE}) (with
$m(E)=const$ and with the safely diagonal $R_{\alpha, \beta}$)
where the evaluation of the sequence of the non-vanishing overlaps
may be reduced to the formula
 \be
 R^{(harm. oscil.)}_{\alpha, \alpha} \sim \sum_{n=0}^{N(\alpha)}
  c_n\,\int_0^\infty
  e^{-x^2}
  \,x^{const+2n}
 dx
 \label{nenum}
 \ee
which evaluates to a sum of $\Gamma-$functions in both the
Hermitian \cite{Bohr} and non-Hermitian \cite{Quesne} cases.
Difficulties with the determination of $R$ perceivably increase
whenever one tries to move to a more realistic model. {\it Vice
versa}, the requirement of the preservation of a manifestly
non-numerical form of the overlaps $R$ leads directly to one of
the eligible definitions of the concept of the so called
quasi-exact (QE) solvability \cite{Ushveridze}. For an explicit
illustration of this statement, let us pick up one of the most
popular QE models, viz., the sextic QE anharmonic-oscillator
generalization of eq. (\ref{hSE}) which, in the units $\hbar^2=
2m(E)=C=1$, reads
 \be
 \left [
 - \frac{\rm d^2}{\rm dr^2} +
 A(E_{\alpha})\,r^2 +
 B\,r^4 +
 C\,r^6\,
 \right ]
 \Psi_{\alpha}({r})= E_{\alpha} \Psi_{\alpha}({r})\,.
 \label{seSE}\
 \ee
This QE model was discovered by Singh et al as early as in the
late seventies \cite{Singh} but its analytic continuation to a
non-Hermitian regime appeared only very recently \cite{Cannata}.
In both these cases we may use our above composite-index notation
with $\alpha = (N,j)$ and $j=1,2,\ldots,m(N)$ and emphasize that
the bound-state solutions of eq. (\ref{seSE}) remain
``quasi-exact" (i.e., elementary and proportional to polynomials)
if and only if we keep the integer $N$ fixed. We must select an
appropriate specific ``spring constant"
$A(E_{\alpha})=A(E^{(N,j)})\equiv A_N$ which is different for the
different $N= 1,2, \ldots$. In this manner one introduces a
certain ``minimally complicated" energy dependence in the
Hamiltonian (\ref{seSE}). At the same time, all the sums and
integrals which define the diagonal as well as non-diagonal
overlaps $R_{\alpha, \beta}$  remain very similar to eq.
(\ref{nenum}) and may be easily shown to degenerate to finite sums
of incomplete $\Gamma-$functions.

\section{Hermitian Hamiltonians $H(E)$  \label{Hilbert} }

\subsection{Bi-orthogonality and completeness}

In eq. (\ref{overlaps}), a nontrivial assumption is needed as a
guarantee that the matrix $R$ has an inverse (for the time being,
let us also keep in mind that it is, by construction and/or
assumption, Hermitian). Only then, the formal decomposition of the
identity projector becomes available as a double sum over the
suitable (sub)set $A$ of the multi-indices,
 \be
\hat{I} = \sum_{\alpha, \beta\in A}\, |\phi_\alpha \rangle \left (
R^{-1} \right )_{\alpha, \beta}
 \langle \phi_\beta|\,.
 \label{CR}
 \ee
Even if we knew the set $A$ and proved the convergence, the
practical value of such a formula would remain significantly
lowered by the non-diagonality of the infinite-dimensional matrix
$R$. The sufficiently precise evaluation of this matrix {\em and}
of its inverse $R^{-1}$ is needed. As we already mentioned (cf.
also \cite{Montreal}), all this requires a sufficiently elementary
form of the basis states $|\phi_\alpha \rangle$.

Equation (\ref{CR}) acquires the standard linear-algebraic meaning
of a ``completeness" relation for our (non-orthogonal) basis
$\{\,|\phi_\alpha \rangle\,\}_{\alpha \in A}$. After we abbreviate
 \ben
\langle\langle \phi_\alpha| = \sum_{ \beta\in A}\, \left ( R^{-1}
\right )_{\alpha, \beta}
 \langle \phi_\beta|\,
 \label{ufe}
 \een
a new basis emerges which, by construction, obeys the
Kronecker-delta-overlap rule
 \ben
\langle \langle \phi_\alpha |\phi_\beta \rangle = \langle
\phi_\alpha |\phi_\beta \rangle\rangle =\delta_{\alpha,
 \beta}, \ \ \ \ \ \ \alpha, \beta \in A\,
 \een
and is called, on this ground, bi-orthogonal and bi-orthonormal
\cite{Iochvidov}.

\subsection{Two alternative quasi-Hermitian representants}

Practical use of the bi-orthogonally generalized basis exhibits a
lot of parallels with its ordinary orthogonal predecessor. First
of all, it enables us to re-interpret the double series (\ref{CR})
as a single-index expansion employing the two different types of
vectors,
 \ben
\hat{I} = \sum_{\alpha\in A}\, |\phi_\alpha \rangle \langle
 \langle \phi_\alpha|=
  \sum_{\alpha\in A}\, |\phi_\alpha \rangle \rangle
 \langle \phi_\alpha|\,.
 \label{CRA}
 \een
It is easy to verify that the new operator defined by its
generalized spectral representation
 \ben
K = \sum_{\alpha, \beta \in A}\, |\phi_\alpha \rangle \,E_\alpha\,
\left ( R^{-1} \right )_{\alpha, \beta} \langle \phi_\beta| \,
 \label{Ka}
 \een
or, in an abbreviated notation,
 \ben
K = \sum_{\alpha \in A}\, |\phi_\alpha \rangle \,E_\alpha\,\langle
 \langle \phi_\alpha|
\,
 \een
may be re-interpreted as an operator which shares with $H(E)$ its
action to the right,
 \ben
 K\,|\phi_\alpha\rangle =E_\alpha\,|\phi_\alpha\rangle\,, \ \ \ \ \
 \ \alpha \in A\,.
 \label{SEK}
 \een
By construction, the new operator $K$ is energy-independent (i.e.,
no nonlinearity is encountered). The latter property is
counterbalanced by non-Hermiticity $K \neq K^\dagger$ which means
that the action of $K$ to the left is much more complicated.
Nevertheless, we may complement the linear, non-Hermitian operator
$K$ by another operator, viz., by
 \ben
L = \sum_{\alpha, \beta \in A}\, |\phi_\alpha \rangle\, \left (
R^{-1} \right )_{\alpha, \beta} \,E_\beta\, \langle \phi_\beta|
\,.
 \label{La}
 \een
This means that
 \ben
 \langle \phi_\alpha|\,L=E_\alpha\,\langle \phi_\alpha|\,, \ \ \ \
 \ \ \ \ \ \  L = \sum_{\beta \in A}\, |\phi_\beta \rangle
 \rangle\,E_\beta\,
 \langle \phi_\beta|
 \,
 \label{ref}
 \een
so that the second non-Hermitian linear operator $L$ shares with
$H(E)$ its action to the left. In this sense, the single nonlinear
operator $H(E)$ may be understood as represented by the linear
doublet $(K,L)$.

One has to notice that $K=L^\dagger$ and $L=K^\dagger$ are
mutually inter-related,
 \be
 \xi\, L = K\,\xi=
  \sum_{\alpha \in A}\, |\phi_\alpha \rangle
  \,E_\alpha\,
 \langle \phi_\alpha|, \ \ \ \ \ \ \ \ \ \
 \xi =
  \sum_{\alpha \in A}\, |\phi_\alpha \rangle
  \,
 \langle \phi_\alpha| = \xi^\dagger\,
 \label{referge}
 \ee
or, alternatively,
 \be
 L \xi^{-1} = \xi^{-1}\,K=
  \sum_{\alpha \in A}\, |\phi_\alpha \rangle  \rangle
  \,E_\alpha\,\langle
 \langle \phi_\alpha|, \ \ \ \ \ \ \ \ \ \
 \xi^{-1}=
  \sum_{\alpha \in A}\, |\phi_\alpha \rangle  \rangle
  \,\langle
 \langle \phi_\alpha| \,.
 \label{referwu}
 \ee
As long as we have $\xi > 0$, the terminology of the review paper
\cite{Geyer} may be used implying that both our mutually conjugate
Hamilton-like operators $K$ and $L$ are quasi-Hermitian. For this
reason, each of them admits a consistent quantum-mechanical
interpretation (readers are recommended to check ref. \cite{Geyer}
for more details).

\section{Non-Hermitian Hamiltonians $H(E)$  \label{postHilbert}}

Recently, several groups of authors \cite{BB} - \cite{Weigert}
tried to weaken the traditional Hermiticity $H = H^\dagger$
[tacitly also assumed, up to now, in formulae (\ref{SE1}) and
(\ref{SE2}) above]. The standard pattern of such a generalization
(which, for the sake of simplicity, keeps the energies real) lies
in the assumption that the bra and ket vectors in the similar
spectral expansions are {\em not} the mere Hermitian conjugates of
each other. This resembles the above-described relation between
the vectors $|\phi_\alpha\rangle$ and $|\phi_\alpha\rangle\rangle$
which we also assumed to remain significantly different. Indeed,
many rules of our preceding section \ref{Hilbert} will find their
parallels also in the forthcoming text where we shall admit that
$H(E) \neq H^\dagger (E)$.

\subsection{Quasi-Hermitian input Hamiltonians $H(E)$ }

Let us keep the {\em superscripted} kets $|\Psi^{(n)}\rangle $ of
eq. (\ref{SE1}) unchanged and introduce another, independent
infinite series of their {\em subscripted} partners
$|\Psi_{(n)}\rangle \neq |\Psi^{(n)}\rangle $. In the light of the
current textbooks \cite{Iochvidov} the usual orthogonality (and
normalization) assumption $\langle \Psi^{(m)} |\Psi^{(n)}\rangle =
\delta_{m,n}$ finds its most natural generalization in the so
called bi-orthogonality (or rather bi-orthonormality) relations
between these two sets,
 \ben
 \langle \Psi_{(m)} |\Psi^{(n)}\rangle = \delta_{m,n},
 \ \ \ \ \ \ \ m, n = 1, 2, \ldots \, .
 \een
In parallel, the usual completeness relations for a basis in our
Hilbert space ${\cal H}$ must be replaced by the following
innovated, bi-orthogonal-basis-related formula
 \be
\hat{I} = \sum_{n}\, |\Psi^{(n)}\rangle
   \langle \Psi_{(n)}|\,.
 \label{CoR}
 \ee
This enables us to replace the spectral decomposition (\ref{SE1})
of Hermitian $H=H^\dagger$ by its non-Hermitian
bi-orthogonal-basis analogue or generalization,
 \be
 H= \sum_n\,|\Psi^{(n)}\rangle \,E^{(n)}\,\langle\Psi_{(n)}|
  \neq H^\dagger\,.
 \label{SE11}
 \ee
Parallels are preserved with eqs. (\ref{referwu}) and/or
(\ref{referge}),
 \be
 H^\dagger \eta = \eta\,H=
  \sum_{n}\, |\Psi_{(n)} \rangle
  \,E^{(n)}\,\langle
 \langle \Psi_{(n)}|, \ \ \ \ \ \ \ \ \ \
 \eta=
  \sum_{n}\, |\Psi_{(n)} \rangle
  \,\langle
 \Psi_{(n)}| = \eta^\dagger\,,
 \label{referli}
 \ee
 \be
 H \eta^{-1} = \eta^{-1}\,H^\dagger=
  \sum_{n}\, |\Psi^{(n)} \rangle
  \,E^{(n)}\,\langle
 \langle \Psi^{(n)}|, \ \ \ \ \ \ \ \ \ \
 \eta^{-1}=
  \sum_{n}\, |\Psi^{(n)} \rangle
  \,\langle
 \Psi^{(n)}| \,.
 \label{referpo}
 \ee
We may call again {\em all} our generalized Hamiltonians
(\ref{SE11}) quasi-Hermitian \cite{Geyer} since we are working
with the manifestly regular and positive $\eta > 0$ in the latter
two relations.

\subsection{Innovated pair of the representants $K$ and $L$}

Once we keep in mind the quasi-Hermiticity symmetries
(\ref{referli}) and (\ref{referpo}), we may introduce an
error-checking convention under which we only consider the
formulae where all the kets are upper-indexed while all the bras
are lower-indexed. Under this convention the extension of the
results of section \ref{Hilbert} to all the quasi-Hermitian $H(E)
\neq H^\dagger(E)$ is straightforward. Firstly, in the way
inspired by eq. (\ref{SE2}) we have to define
 \be
 H_{}(z)= \sum_n\,|\Psi^{(n)}_{}(z)\rangle \,E^{(n)}_{}(z)
 \,\langle\Psi_{(n)}(z)|\,
 \label{SE22}
 \ee
and re-abbreviate
 \be
 |\varphi^\alpha \rangle=
 |\Psi^{(n)}_{}(z_{phys})\rangle, \ \ \ \ \ \
 \langle
 \varphi_\alpha|=\langle\Psi_{(n)}(z_{phys})|\,
 \ee
where the rising or lowering of the index means a transition to
another, {\em entirely different} vector. Assuming the knowledge
of all the overlaps
 \ben
 \langle \varphi_\alpha |\varphi^\beta \rangle =R_{\alpha}^{\ \beta}
 \,
 \label{overlapspt}
 \een
(notice an inessential change in our matrix indexing), we have to
emphasize that they do not form a Hermitian matrix anymore. Still,
its assumed invertibility suffices for us to write down the unit
projector
 \be
\hat{I} = \sum_{\alpha, \beta\in A}\, |\varphi^\alpha \rangle
\left ( R^{-1} \right )_{\alpha}^{\ \beta}
 \langle \varphi_\beta|
 =
 \sum_{\beta\in A}\, |\varphi^\beta \rangle
 \rangle \,\langle \varphi_\beta|
 =
 \sum_{\alpha\in A}\, |\varphi^\alpha \rangle\,
 \langle \langle \varphi_\alpha|
 \,
 \label{CRpo}
 \ee
where we abbreviated
 \ben
 \sum_{\alpha\in A}\, |\varphi^\alpha \rangle \left (
 R^{-1} \right )_{\alpha}^{\ \beta}
 \, \equiv\,
 |\varphi^\beta\rangle \rangle  \,,
 \ \ \ \ \ \ \ \ \ \ \ \ \
 \sum_{ \beta\in A}\,
 \left ( R^{-1} \right )_{\alpha}^{\ \beta}
 \langle \varphi_\beta| \, \equiv\,
 \langle \langle \varphi_\alpha| \,.
 \label{ufept}
 \een
This means that
 \ben
\langle \langle \varphi_\alpha |\varphi^\beta \rangle = \langle
\varphi_\alpha |\varphi^\beta \rangle\rangle =\delta_{\alpha}^{\
\beta}, \ \ \ \ \ \ \alpha, \beta \in A\,
 \een
and that
 \ben
\hat{I} = \sum_{\alpha\in A}\, |\varphi^\alpha \rangle \langle
 \langle \varphi_\alpha|=
  \sum_{\alpha\in A}\, |\varphi^\alpha \rangle \rangle
 \langle \varphi_\alpha|\,.
  \een
It is now trivial to define the generalized operators $K$ and $L$
which share with the quasi-Hermitian $H(E)$ its action to the
right and left, respectively,
 \ben
K = \sum_{\alpha \in A}\, |\varphi^\alpha \rangle
\,E_\alpha\,\langle
 \langle \varphi_\alpha|
\,\ \ \ \ \ \
 K\,|\varphi^\beta\rangle =E_\beta\,|\varphi^\beta\rangle\,,
 \een
 \ben
 \langle \varphi_\alpha|\,L=E_\alpha\,\langle \varphi_\alpha|\,, \ \ \ \
 \ \ \ \ \ \  L = \sum_{\beta \in A}\, |\varphi^\beta \rangle
 \rangle\,E_\beta\,
 \langle \varphi_\beta|
 \,.
 \een
This completes our construction.

\section{Summary: Separation of the left and right action of the
 energy-dependent Hamiltonians in Hilbert space}

When we compare the results of sections \ref{Hilbert} and
\ref{postHilbert} we may feel surprised by the smoothness of the
transition to the non-Hermitian $H(E)$. In fact, the only
perceivable consequence of the loss of the Hermiticity of $H(E)$
lies in the emergence of the independent operator of the inverted
metric which is defined in terms of the new, upper-indexed bras
and kets. Formally, the breakdown $K \neq L^\dagger$ and $L \neq
K^\dagger$ of the conjugation symmetry only implies that the
quasi-Hermiticity rules (\ref{referli}) and (\ref{referpo}) must
be replaced by their appropriate modifications
 \ben
 K^\dagger \mu = \mu\,K=
  \sum_{\alpha \in A}\, |\varphi_\alpha \rangle  \rangle
  \,E_\alpha\,\langle
 \langle \varphi_\alpha|, \ \ \ \ \ \ \ \ \ \
 \mu=
  \sum_{\alpha \in A}\, |\varphi_\alpha \rangle  \rangle
  \,\langle
 \langle \varphi_\alpha| = \mu^\dagger\,
 \een
and
 \ben
 \nu\, L = L^\dagger\,\nu=
  \sum_{\alpha \in A}\, |\varphi_\alpha \rangle
  \,E_\alpha\,
 \langle \varphi_\alpha|, \ \ \ \ \ \ \ \ \ \
 \nu =
  \sum_{\alpha \in A}\, |\varphi_\alpha \rangle
  \,
 \langle \varphi_\alpha| = \nu^\dagger\,
 \label{referwil}
 \een
where we must keep in mind that
 \be
 \mu^{-1}=
  \sum_{\alpha \in A}\, |\varphi^\alpha \rangle
  \,
 \langle \varphi^\alpha| \,, \ \ \ \ \ \ \ \ \ \
 \nu^{-1} =
  \sum_{\alpha \in A}\, |\varphi^\alpha \rangle \rangle
  \,
 \langle \langle \varphi^\alpha| \,.
 \label{referwsa}
 \ee
This means that we in fact violated the ``error-correcting"
convention of section \ref{postHilbert} and started writing all
the ``redundant" Hermitian-conjugate formulae in their explicit
form. Such a step facilitates the most concise formulation of our
present message stating that the original operator $H(E)$ and its
conjugate $H^\dagger(E)$ are in fact {\em subtly} different in
{\em both} the Hermitian and non-Hermitian cases. In the other
words, even after the present ``linearization" of their action in
the standard, self-dual Hilbert space, {\em none} of their two
energy-independent representatives $K$ and $L$ is redundant.

\section*{Acknowledgements}

Work supported by  GA AS {C}R, grant Nr. 104 8302
 and by
the AS CR projects K1048102, K1010104 and AV0Z1048901.

\end{document}